%
%
%
%
%
%
%
%
%
%
%
%
%
%
%
%
%
\input phyzzx
%
%
\catcode`\@=11 
\def\papersize{\hsize=40pc \vsize=53pc \hoffset=0pc \voffset=1pc
   \advance\hoffset by\HOFFSET \advance\voffset by\VOFFSET
   \pagebottomfiller=0pc
   \skip\footins=\bigskipamount \normalspace }
\catcode`\@=12 
\papers
\def\to{\rightarrow}
\def\half{\textstyle{1\over 2}}

\vsize=23.cm
\hsize=15.cm

\tolerance=500000
\overfullrule=0pt

\Pubnum={LPTENS-95/53 \cr
{\tt hep-th@xxx/9601007} \cr
December 1995}

\date={}
\pubtype={}
\titlepage
\title{\bf DUALITY IN $N=2$ SUSY $SU(2)$ YANG-MILLS THEORY:\break
A pedagogical introduction to the work of Seiberg and Witten}  
\author{Adel~Bilal}
\address{
CNRS - Laboratoire de Physique Th\'eorique de l'Ecole
Normale Sup\'erieure
\foot{{\rm unit\'e propre du CNRS, associ\'e \`a l'Ecole Normale
Sup\'erieure et l'Universit\'e Paris-Sud}}
 \nextline 24 rue Lhomond, 75231
Paris Cedex 05, France\break
{\tt bilal@physique.ens.fr}
}

\abstract{These are notes from introductory lectures given at the Ecole
Normale in Paris and at the Strasbourg meeting dedicated to the memory of
Claude Itzykson.\nextline
\indent 
I review in considerable detail and in a hopefully pedagogical way the
work of Seiberg and Witten on $N=2$ supersymmetric $SU(2)$ gauge theory without
extra matter. This presentation basically follows their original work, except
in the last section where the low-energy effective action is
obtained emphasizing more the relation between monodromies and differential
equations rather than using elliptic curves.}

\vskip 1.cm
\centerline{\it To appear in the Proceedings of the ``61. Rencontre entre
Physiciens Th\'eoriciens } 
\centerline{\it et Math\'ematiciens", Strasbourg, France, December 1995}
\centerline{\it dedicated to the memory of Claude Itzykson}

\endpage
\pagenumber=1

 \def\PL #1 #2 #3 {Phys.~Lett.~{\bf #1} (#2) #3}
 \def\NP #1 #2 #3 {Nucl.~Phys.~{\bf #1} (#2) #3}
 \def\PR #1 #2 #3 {Phys.~Rev.~{\bf #1} (#2) #3}
 \def\PRL #1 #2 #3 {Phys.~Rev.~Lett.~{\bf #1} (#2) #3}
 \def\CMP #1 #2 #3 {Comm.~Math.~Phys.~{\bf #1} (#2) #3}
 \def\IJMP #1 #2 #3 {Int.~J.~Mod.~Phys.~{\bf #1} (#2) #3}
 \def\JETP #1 #2 #3 {Sov.~Phys.~JETP.~{\bf #1} (#2) #3}
 \def\PRS #1 #2 #3 {Proc.~Roy.~Soc.~{\bf #1} (#2) #3}
 \def\IM #1 #2 #3 {Inv.~Math.~{\bf #1} (#2) #3}
 \def\JFA #1 #2 #3 {J.~Funkt.~Anal.~{\bf #1} (#2) #3}
 \def\LMP #1 #2 #3 {Lett.~Math.~Phys.~{\bf #1} (#2) #3}
 \def\IJMP #1 #2 #3 {Int.~J.~Mod.~Phys.~{\bf #1} (#2) #3}
 \def\FAA #1 #2 #3 {Funct.~Anal.~Appl.~{\bf #1} (#2) #3}
 \def\AP #1 #2 #3 {Ann.~Phys.~{\bf #1} (#2) #3}
 \def\MPL #1 #2 #3 {Mod.~Phys.~Lett.~{\bf #1} (#2) #3}

\def\a{\alpha}
\def\adot{{\dot\alpha}}
\def\b{\beta}
\def\m{\mu}
\def\s{\sigma}
\def\n{\nu}

\def\l{\lambda}
\def\lb{{\bar\lambda}}
\def\L{\Lambda}
\def\g{\gamma}
\def\G{\Gamma}
\def\t{\theta}
\def\tb{{\bar\theta}}
\def\sm{\sigma^\m}
\def\smb{\bar\sigma^\m}
\def\d{\partial}
\def\rmd{{\rm d}}
\def\vf{\varphi}
\def\f{\phi}
\def\F{\Phi}
\def\fd{\phi_D}
\def\Fd{\Phi_D}
\def\ix{\int {\rm d}^4 x \, }
\def\dt{{\rm d}^2 \t \, }
\def\dtb{{\rm d}^2 \tb \, }
\def\dtt{{\rm d}^2 {\tilde\t} \, }
\def\P{\Psi}
\def\cf{{\cal F}}
\def\cfd{{\cal F}_D}
\def\ad{a_D}
\def\e{\epsilon}
\def\rmd{{\rm d}}
\def\rd{\sqrt{2}}
\def\la{\langle}
\def\ra{\rangle}

\def\P{\Psi}
\def\p{\psi}
\def\pb{{\bar \psi}}
\def\dd #1 #2{{\delta #1\over \delta #2}}
\def\tr{{\rm tr}\ }

\def\nm{\nabla_\m}
\def\fmn{F_{\m\n}}
\def\fmnt{{\tilde F}_{\m\n}}
\def\im{{\rm Im}\, }
\def\M{{\cal M}}

\REF\SW{N. Seiberg and E. Witten, {\it Electric-magnetic duality, monopole
condensation, and confinement in $N=2$ supersymmetric Yang-Mills theory}, 
\NP B426 1994 19 , {\tt hep-th/9407087}.}
\REF\LER{A. Klemm, W. Lerche, S. Yankielowicz and S. Theisen, {\it Simple singularities
and $N=2$ supersymmetric Yang-Mills theory},  \PL B344 1995 169 , {\tt
hep-th/9411048};\nextline
A. Klemm, W. Lerche and S. Theisen, {\it Nonperturbative effective actions of
$N=2$ supersymmetric gauge theories}, CERN preprint CERN-TH/95-104, {\tt
hep-th/9505150}.}
\REF\OGG{P.C. Argyres and A.E. Faraggi, {\it The vacuum structure and spectrum of $N=2$
supersymmetric $SU(N)$ gauge theory}, \PRL 74 1995 3931 , {\tt hep-th/9411057};\nextline
U.H. Danielsson and B. Sundborg, {\it The moduli space and monodromies of $N=2$
supersymmetric $SO(2r+1)$ Yang-Mills theory}, \PL B358 1995 273 , 
{\tt hep-th/9504102};\nextline
A. Brandhuber and K. Landsteiner, {\it On the monodromies of $N=2$ supersymmetric
Yang-Mills theory with gauge group $SO(2N)$}, \PL B358 1995 73 , 
{\tt hep-th/9507008}. }
\REF\MAT{N. Seiberg and E. Witten, {\it Monopoles, duality and chiral symmetry breaking
in $N=2$ supersymmetric QCD}, \NP B431 1994 484 , {\tt hep-th/9408099};\nextline
P.C. Argyres and A.D. Shapere, {\it The vacuum structure of $N=2$ super QCD with
classical gauge groups}, Rutgers preprint RU-95-61, {\tt hep-th/9509175}.}
\REF\POL{J. Polchinski, {\it Recent results in string duality}, ITP Santa Barbara
preprint, {\tt hep-th/9511157}.}
\REF\STRINGS{A. Strominger, {\it Black hole condensation and duality in string theory},
Santa Barbara preprint UCSBTH-95-29, {\tt hep-th/9510207}. }
\REF\ALG{C. Gomez, {Electric-magnetic duality and effective field theories}, Madrid
preprint FTUAM 95/36, {\tt hep-th/9510023}; \nextline
L. Alvarez-Gaum\'e, notes of lectures given at CERN in the fall of 1994, to be
published.}
\REF\THO{G. 't Hooft, \NP B79 1974 276 ;   \nextline
 A.M. Polyakov,  JETP Letters {\bf 20} (1974) 194.}
\REF\MO{C. Montonen and D. Olive, {\it Magnetic monopoles as gauge particles?}, 
\PL 72B 1977 117 .}
\REF\OLI{D. Olive, {\it Exact electromagnetic duality}, Swansea preprint SWAT/94-95/81,
{\tt hep-th/9508089}.}
\REF\SUSY{M.F. Sohnius, {\it Introducing supersymmetry}, 
Phys. Rep. {\bf 128} (1985) 39;\nextline
P. Fayet and S. Ferrara, {\it Supersymmetry}, Phys. Rep. {\bf 32C} (1977) 1.}
\REF\SEI{N. Seiberg, {\it Supersymmetry and non-perturbative beta functions},
\PL B206 1988 75 .}
\REF\COL{S. Coleman, {\it Classical lumps and their quantum descendants}, in:
Aspects of Symmetry, Cambridge University Press, 1985}
\REF\OW{E. Witten and D. Olive, {\it Supersymmetry algebras that include topological
charges}, \PL 78B 1978 97 .}
\REF\SEN{A. Sen, {\it Dyon-monopole bound states, self-dual harmonic forms on the
multi-monopole moduli space, and $Sl(2,{\bf Z})$ invariance in string theory}, \PL 329B
1994 217 , {\tt hep-th/9402032}.}
\REF\ERD{A. Erdelyi et al, {\it Higher Transcendental Functions}, Vol 1, McGraw-Hill,
New York, 1953.}
\REF\BG{A. Bilal and J.-L. Gervais, {\it Exact quantum three-point function of
Liouville highest weight states}, \NP B305 1988 33 .}
\REF\DS{U.H. Danielsson and B. Sundborg, {\it Exceptional Equivalences in $N=2$
supersymmetric Yang-Mills theory}, Uppsala preprint USITP-95-12, {\tt
hep-th/9511180}.}
\REF\MVL{S. Kachru, A. Klemm, W. Lerche, P. Mayr and C. Vafa, 
{\it Nonperturbative results on the point particle limit of N=2 heterotic
 string compactifications}, CERN/Harvard preprint CERN-TH/95-231, HUTP-95/A032,
{\tt hep-th/9508155}.}

{\bf \chapter{Introduction}}
\sectionnumber=0

Although a quite old one, the notion of duality has become most central in
field and string theory during the last year and a half. The major
breakthrough in field theory was the paper by Seiberg and Witten [\SW]
considering the pure $N=2$ supersymmetric $SU(2)$ Yang-Mills theory. This work
was then generalized to other gauge groups [\LER,\OGG] and to theories including extra
matter fields [\MAT]. In the same time, it became increasingly clear that
dualities in string theories play a maybe even more fascinating role (for a
brief review see e.g. [\POL] or [\STRINGS]). Rather than attempting to give an
overview of the situation, in the present notes I will try to give a {\it
pedagogical} introduction to the first paper by Seiberg and Witten [\SW].
Several other introductions do exist [\ALG], and I hope that the present
notes complement them in a useful way.

The idea of duality probably goes back to Dirac who observed that the
source-free Maxwell equations are symmetric under the exchange of the
electric and magnetic fields. More precisely, the symmetry is $E\to B,\ B\to
-E$, or $\fmn\to \fmnt=\half
\e_{\m\n}^{\phantom{\m\n}\rho\sigma}F_{\rho\sigma}$. (Here
$\e_{\m\n\rho\sigma}$ is the flat-space antisymmetric $\e$-tensor with
$\e^{0123}=+1$ and $\eta_{\m\n}$ has signature $(1,-1,-1,-1)$.) To maintain
this symmetry in the presence of sources, Dirac introduced, somewhat ad hoc,
magnetic monopoles with magnetic charges $q_m$ in addition to the electric
charges $q_e$, and showed that consistency of the quantum theory requires a
charge quantization condition $q_mq_e=2\pi n$ with integer $n$. Hence the
minimal charges obey $q_m={2\pi\over q_e}$. Duality exchanges $q_e$ and
$q_m$, i.e. $q_e$ and ${2\pi\over q_e}$. Now recall that the electric charge
$q_e$ also is the coupling constant. So duality exchanges the coupling
constant with its inverse (up to the factor of $2\pi$), hence exchanging
strong and weak coupling. This is the reason why we are so much interested in
duality: the hope is to learn about strong-coupling physics from the
weak-coupling physics of a dual formulation of the theory. Of course, in
classical Maxwell theory we know all we may want to know, but this is no
longer true in quantum electrodynamics.

Actually, quantum electrodynamics is not a good candidate for exhibiting a
duality symmetry since there are no magnetic monopoles, but the latter
naturally appear in spontaneously broken non-abelian gauge theories
[\THO]. Unfortunately, electric-magnetic duality in its simplest form cannot
be a symmetry of the quantum theory due to the running of the coupling
constant (among other reasons). Indeed, if duality exchanges $\a(\L)
\leftrightarrow {1\over\a(\L)}$ (where $\a(\L)={4\pi\over e^2(\L)}$) at some
scale $\L$, in general this won't be true at another scale. This argument is avoided if
the coupling does not run, i.e. if the $\b$-function vanishes as is the case
in certain ($N=4$) supersymmetric extensions of the Yang-Mills theory. This
and other reasons led Montonen and Olive [\MO] to conjecture that duality
might be an exact symmetry of $N=4$ susy Yang-Mills theory. A nice review of these
ideas can
be found in [\OLI].

Let me recall that a somewhat similar duality symmetry appears in the two-dimensional
Ising model where it exchanges the temperature with a dual temperature, thereby
exchanging high and low temperature analogous to strong and weak coupling.
For the Ising model, the sole existence of the duality symmetry led to the 
exact determination of the critical temperature as the self-dual point, well prior to
the exact solution by Onsager. One may
view the existence of this self-dual point as the requirement that the dual
high and low temperature regimes can be consistently ``glued" together.
Similarly, in the Seiberg-Witten theory, as will be explained below, duality
allows us to obtain the full effective action for the light fields at any
coupling (the analogue of the Ising free energy at any temperature) from
knowledge of its weak-coupling limit and the behaviour at certain
strong-coupling ``singularities", together with a holomorphicity requirement that tells
us how to patch together the different limiting regimes.

Let me give an overview of how I will proceed. $N=2$ supersymmetry is central
to the work of Seiberg and Witten and to the way duality works, so we must
spend some time in the next section to review those notions of supersymmetry
that we will need, including the formulation of the $N=2$ super Yang-Mills
action. In section 3, I will discuss the Wilsonian low-energy effective
action corresponding to the (microscopic) $N=2$ super Yang-Mills action for the gauge
group $SU(2)$. The original $SU(2)$ gauge symmetry has been broken down to
$U(1)$ by the expectation value $a$ of the scalar field $\f$ contained in the
$N=2$ multiplet, and the effective action describes the physics of the
remaining massless $U(1)$ susy multiplet in terms of an a priori {\it
unknown} function $\cf(a)$. $N=2$ supersymmetry constrains $\cf$ to be a
(possibly multivalued) {\it holomorphic} function. Different vacuum expectation
values $a$, or rather different values of the gauge-invariant vacuum expectation
value $u=\la\tr \f^2\ra$ lead to physically different theories. So $u$
parametrizes the space of inequivalent vacua, called the moduli
space.

In section 4, I will discuss how one defines the duality transformations and
show that duality inverts a certain combination $\tau$ of the effective coupling
constant and the effective theta angle. I will also discuss the spectrum of massive
states (BPS mass formula). Let me insist that this duality is an {\it exact}  
symmetry of the
abelian low-energy {\it effective} theory, not of the microscopic $SU(2)$ theory. 
This is
different from the Montonen-Olive conjecture about an exact duality symmetry of a
microscopic gauge theory.

In section 5, we will study the behaviour of
the low-energy effective theory at certain singular points in the moduli
space, i.e. at certain values of the parameter $u$ where a magnetic monopole or a dyon
(an electrically and magnetically charged state) becomes massless, leading to
a singularity of the effective action. These singularities translate into
certain monodromies of $a\sim \la\f\ra$ and its dual partner
$\ad={\d\cf(a)\over \d a}$. In section 6, we put everything together, and I
show how to obtain $a(u)$ and $\ad(u)$ and hence $\cf(a)$ from the knowledge of these
monodromies. Then the low energy effective action is known and the theory
solved for all values of the effective coupling constant $\tau=
{\d^2\cf(a)\over \d a^2}$. Section 6 is the only part where the presentation
does not follow the logic of Seiberg and Witten's paper, but I rather
emphasize the relation between monodromies and differential equations, and
obtain $a(u)$ and $\ad(u)$ as solutions of a hypergeometric equation. I then
show how this fits into the reasoning of Seiberg and Witten using elliptic
curves. In a concluding section 7, I mention some of the developments that
followed the work of Seiberg and Witten described in these notes.

{\bf \chapter{Some notions of supersymmetry}}
\sectionnumber=0

Clearly, I cannot give a complete discussion of the theory of $N=2$
supersymmetry, see e.g. ref. [\SUSY]. Instead, I will introduce just as much
as I believe is necessary to understand the basic features of the $N=2$
supersymmetric Yang-Mills theory. Maybe I should stress that all of the 
physics will be in four dimensional Minkowski space, so the supersymmetries
all refer to the standard $D=4$ case. A standard Dirac spinor then has four
complex components and transforms reducibly under the action of the (covering
group $Sl(2,{\bf C})$ of the) Lorentz group. It is more convenient to break
such a Dirac spinor into pieces each having 2 complex components and
transforming irreducibly. These two-component spinors are denoted $\chi^\a$ and
$\bar\chi_\adot =(\chi^\a)^*$ according to their Lorentz transformation
properties. Dealing with two-component spinors, one also encounters the
matrices $\sm$ and $\smb$: $(\sm)_{\a\adot}$ being the unit matrix for $\m=0$
and the Pauli matrices $\sigma^i$ for $\m=i=1,2,3$, while for $\smb$ one has
$-\sigma^i$ instead. For completeness, I mention that one also needs the
antisymmetric tensor $\e^{\a\b}$ with $\e^{01}=+1$ and its inverse to raise
and lower spinor indices. The convention for contracting indices is
$\p\chi\equiv\p^\a\chi_\a$ and $\bar\p\bar\chi=\bar\p_\adot \bar\chi^\adot$.

\section{Unextended supersymmetry}

The simplest, unextended supersymmetry (called $N=1$ susy in contrast with
the extended $N>1$ susys) can be represented on a variety of multiplets of
fields involving bosons and fermions. One of the simplest representations
involves a complex scalar field $\f$ and a two-component spinor $\p_\a$
($\a=1,2$). They form the so-called chiral scalar multiplet. I do not
write the susy transformations since we do not need them here (see [\SUSY]).
To write down susy invariant Lagrangians (actions) it is convenient to
assemble $\f$ and $\p$ into a superfield. Therefore one introduces
anticommuting varaiables $\t^\a$ and $\tb_\adot$ and writes
$$\F=\f(y)+\rd \t\p(y)+\t^2 F(y)
\eqn\di$$
where $y^\m=x^\m+i\t\sm\tb$ and $\t\p\equiv\t^\a\p_\a$, $\t^2\equiv
\t\t\equiv\t^\a\t_\a=-2\t^1\t^2$, $\t\sm\tb\equiv
\t^\a\sm_{\a\adot}\tb^\adot$. (Notice that $\t^2$ is used to denote $\t\t$ 
as well as the second component of $\t$. It should be clear which one is
meant, and almost always it is $\t\t$.) $\F$ is a chiral superfield. One also
needed to include a field $F$ that will turn out to be an auxiliary field.
Expanding the $y$-dependence (and using $\t^1\t^1=\t^2\t^2=0$) one finds
$$\eqalign{
\F=&\f(x)+i\t\sm\tb \d_\m\f(x)-{1\over 4}\t^2\tb^2\d^2\f(x)+\rd\t\p(x)\cr
&-{i\over \rd}\t^2 (\d_\m\p(x)\sm\tb) +\t^2 F(x)\ . \cr}
\eqn\dii$$
A supersymmetry invariant action then is given by the superspace integral
$${1\over 4} \ix\dt\dtb \F^+\F \ .
\eqn\diii$$
The $\t$-integrations are defined such that only the term proportional to
$\t^2\tb^2$ in $\F^+\F$ gives a non-vanishing result. (One has $\int\dt\dtb \t^2
\tb^2 =4$.) Then \diii\ becomes
$$\ix\left(\d_\m\f  \d^\m\f^+ -i\pb\smb\d_\m\p+F^+F\right) \ .
\eqn\div$$
We see that the simple $\F^+\F$-term has produced the standard kinetic terms
for a complex scalar $\f$ and the spinor $\p$. $F$ is an auxiliary field
which can be set equal to zero by its equation of motion. Supersymmetry
invariant interactions can be generated by a superpotential $\ix\left[
\int\dt {\cal W}(\F) + {\rm h.c.}\right]$ where ${\cal W}(\F)$ depends only on $\F$
and not on $\F^+$.

Another supersymmetry multiplet is the vector multiplet that contains a
(massles gauge) vector field $A_\m$ and its superpartner $\l_\a$ (gaugino).
They are combined together with an auxiliary field $D$ into a superfield $V$
as \foot{
Actually the form given here is the one obtained after fixing the Wess-Zumino
gauge in the general real superfield $V$ using $V\to V+\L+\L^+$ where $\L$ is
a chiral superfield.}
$$V=-\t\sm\tb A_\m+i\t^2 (\tb\bar\l) -i\tb^2 (\t\l) +\half\t^2\tb^2 D \ .
\eqn\dv$$
We will be interested in the case of non-abelian gauge symmetry where $A_\m$,
and hence $\l,\bar\l $ and $D$ are in the adjoint representation:
$A_\m=A^a_\m T_a$, $[T_a,T_b]=f_{abc}T_c$, etc. From the superfield $V$ one
defines another (spinorial) superfield $W_\a$ as
$$W=\left(-i\l+\t D-i\s^{\m\n}\t \fmn +\t^2\sm\nm\lb\right)(y)
\eqn\dvi$$
(again, $y^\m=x^\m+i\t\sm\tb$), where $\s^{\m\n}={1\over
4}(\sm\bar\s^\n-\s^\n\smb)$, $\fmn=\d_\m A_\n-\d_\n A_\m-ig [A_\m,A_\n]$,
$\nm\l=\d_\m \l-ig[A_\m,\l]$ and $g$ is the gauge coupling constant.
The corresponding superspace formula is
$$W_\a={1\over 8g} \bar D^2\left( e^{2gV}D_\a e^{-2gV}\right) \ .
\eqn\dvii$$
Here $D_\a$ and $\bar D_\adot$ are
the superspace derivatives $\d/\d\t^\a+i\sm_{\a\adot} \tb^\adot \d_\m$ and 
$-\d/\d\tb^\adot-i\sm_{\a\adot} \t^\a \d_\m$. The supersymmetric Yang-Mills
action then simply is (one has $\int\dt \t^2=-2$)
$$
-{1\over 4}\ix\dt \tr W^\a W_\a =\ix \tr\left[-{1\over4}\fmn F^{\m\n}
+{i\over4}\fmn \tilde F^{\m\n}
-i\l\sm\nm\lb +{1\over 2} D^2\right] \ . 
\eqn\dviii$$
In addition to the standard Yang-Mills term $-{1\over4}\fmn F^{\m\n}$ one has
also generated a term ${i\over4}
\fmn \tilde F^{\m\n}$ which, after
integration, gives the instanton number. It should appear in the action
multiplying the $\t$-parameter (not to be confused with the anticommuting
$\t$-variables of superspace!) and with a real coefficient. Hence if one
introduces the complex coupling constant
$$\tau={\t\over 2\pi}+{4\pi i\over g^2}
\eqn\dix$$
then the following {\it real} action precisely does what one wants:
$$\eqalign{
{1\over 16\pi}\im\left[ \tau \ix\dt \tr W^\a W_\a \right]
=&{1\over g^2}\ix\tr\left[-{1\over4}\fmn F^{\m\n} -i\l\sm\nm\lb +{1\over 2}
D^2\right] \cr
&+{\t\over 32\pi^2}\ix \fmn \tilde F^{\m\n}   \cr}
\eqn\dx$$
with the $F^2$-term and the instanton number conventionally normalized.

The matter field $\F$ can be minimally coupled to the Yang-Mills field by
putting it in some representation of the gauge group, say the adjoint, and
replacing \diii\ and \div\ by
$$\eqalign{
&{1\over 4} \ix\dt\dtb \tr \F^+ e^{-2gV}\F \cr
&=\ix \tr \left( \vert\nm\f\vert^2-i\pb\smb\nm\p+F^+F
-g\f^+[D,\f]-\rd ig\f^+\{\l,\p\}+\rd ig \pb[\lb,\f]\right) \ . \cr}
\eqn\dxi$$
In addition to the appearance of the covariant derivatives $\nm$ we also 
see explicit couplings
between $\f,\ \p$ and $\l,\ D$ as required by supersymmetry.

\section{The $N=2$ super Yang-Mills action}

$N=2$ supersymmetry combines all of the fields $\f,\p$ and $A_\m,\l$ into a
single susy multiplet. Of course, this means that all fields must be in the
same representation of the gauge group as $A_\m$, i.e. in the adjoint
representation. This multiplet contains two spinor fields $\p$ and $\l$ on
equal footing. So the simplest guess for the $N=2$ super Yang-Mills action is
a combination of \dx\ and \dxi\ with relative coefficients such that the two
kinetic terms for $\p$ and $\l$ have the same coefficients. Integrating by
parts one of them, we see that we have to add \dx\ and ${1\over g^2}$ times
\dxi. It is by no means obvious that the resulting sum has $N=2$
supersymmetry, but one can check that it does. Thus the $N=2$ super
Yang-Mills action is
$$\eqalign{
S&= \ix \left[\im \left({\tau\over 16\pi} \dt \tr W^\a W_\a \right)
+{1\over 4g^2} \int\dt\dtb \tr \F^+ e^{-2gV}\F\right]\cr
&=\im\  \tr \ix {\tau\over 16\pi} \left[\int\dt  W^\a W_\a 
+\int\dt\dtb  \F^+ e^{-2gV}\F\right]\ . \cr}
\eqn\dxii$$
Note that a non-trivial superpotential ${\cal W}(\F)$ is not allowed by $N=2$
supersymmetry.

An important point concerns the auxiliary fields  in $S$:
$$S_{\rm aux}={1\over g^2} \ix \tr \left[ {1\over 2} D^2
-g\f^+[D,\f]+F^+F\right] \ .
\eqn\dxiii$$
Solving the auxiliary field equations and inserting the result back into the
action gives
$$S_{\rm aux}=-\ix {1\over 2}\, \tr \left( [\f^+,\f]\right)^2
\eqn\dxiv$$
which shows that the bosonic potential is
$V(\f)={1\over 2}\, \tr \left( [\f^+,\f]\right)^2\ge 0$. As is well known, a
ground state field configuration $\f_0$  with $ V(\f_0) > 0$ does break supersymmetry. In other
words, unbroken susy requires a ground state (vacuum) with 
$ V(\f_0)=0$. Note that this does not imply $ \f_0=0$. A  sufficient
and necessary condition is that 
$\f_0$ and $\f_0^+$ commute.

The $N=2$ supersymmetry of \dxii\ can be rendered manifest by using a $N=2$
superspace notation. I will not go into any details and simply quote some
relevant formulas. In addition to the anticommuting $\t_\a, \tb_\adot$ of
$N=1$ susy, one now needs a second set of anticommuting $\tilde\t_\a,
\bar{\tilde{\t}}_\adot$. One introduces the $N=2$ chiral superfield
$$\P=\F(\tilde y,\t)+\rd \tilde\t^\a W_\a(\tilde y,\t)
+\tilde\t^\a\tilde\t_\a G(\tilde y,\t)
\eqn\dxv$$
where $\tilde y^\m=x^\m+i\t\sm\tb+i\tilde\t\sm\bar{\tilde\t}
=y^\m+i\tilde\t\sm\bar{\tilde\t}$ and
$$G(\tilde y,\t)=-{1\over 2} \int\dtb \left[\F(\tilde y -i\t\sigma\tb,\t,\tb )\right]^+
\exp\left[ -2g V(\tilde y -i\t\sigma\tb,\t,\tb) \right]
\eqn\dxvi$$
with $\F(y,\t)$ and $\F(x,\t,\tb)$ as given in \di\ and \dii\ and $W(y,\t)$ as given in \dvi.
The $\dtb$-integration is meant to be at fixed $\tilde y$. $\P$ is the $N=2$ analogue
of a chiral superfield, subject to the constraint \dxvi\ necessary in order to
eliminate certain unphysical degrees or freedeom. The $N=2$ superspace notation 
``implies"
that the following action is $N=2$ susy invariant:
$$\im\left[ {\tau\over 16\pi} \ix\dt\dtt {1\over 2}\, \tr \P^2\right] \ .
\eqn\dxvii$$
Carrying out the $\dtt$-integration yields precisely the action \dxii.

Note that the integrand in \dxvii\ only depends on $\P$, not on $\P^+$. More generally
one can show that $N=2$ supersymmetry constrains the form of the action to be
$$ {1\over 16\pi} \im  \ix\dt\dtt \cf(\P)
\eqn\dxviii$$
where $\cf$, called the $N=2$ prepotential, depends only on $\P$ and not on $\P^+$.
This is referred to as holomorphy of the prepotential. For the $N=2$ super Yang-Mills
action \dxii\ or \dxvii\ one simply has
$$\cf(\P)\equiv \cf_{\rm class}(\P)={1\over 2}\, \tr \tau \P^2 \ .
\eqn\dxix$$
The quadratic dependence on $\P$ is fixed by renormalisability. Below we will consider
low-energy effective actions. Then the only constraint is $N=2$ susy, translated as
holomorphicity of $\cf$. In $N=1$ superspace language, the general action \dxviii\
reads
$$\eqalign{
 {1\over 16\pi} \im\ix \Big[ &\int\dt \cf_{ab}(\F)W^{a\a}W^b_{\phantom{b}\a}\cr
+&\int\dt\dtb\left(\F^+e^{-2gV}\right)^a \cf_a(\F) \Big] \cr }
\eqn\dxx$$
where $\cf_a(\F)={\d\cf(\F)\over \d\F^a}$, 
$\cf_{ab}(\F)={\d^2\cf(\F)\over \d\F^a  \d\F^b}$ and where $a,b$ are Lie algebra
indices, so that $\F=\F^aT_a$, $W_\a=W^a_\a T_a$ with $\tr T_aT_b=\delta_{ab}$. 
This concludes our quick tour through supersymmetric
Yang-Mills theories.

{\bf \chapter{Low-energy effective action of $N=2$ susy $SU(2)$ Yang-Mills theory}}
\sectionnumber=0

Following Seiberg and Witten [\SW] we want to study and determine the low-energy
effective action of the $N=2$ susy Yang-Mills theory with gauge group $SU(2)$. The
latter theory is the microscopic theory which controls the high-energy behaviour. It
is renormalisable and well-known to be asymptotically free. The low-energy effective
action will turn out to be quite different.

{\section{Low-energy effective actions}}

There are two types of effective actions. One is the standard generating functional
$\Gamma[\vf]$ of one-particle irreducible Feynman diagrams (vertex functions). It is
obtained from the standard renormalised generating functional $W[\vf]$ of connected
diagrams by a Legendre transformation. Momentum integrations in loop-diagrams are from
zero up to a UV-cutoff which is taken to infinity at the end.
$\Gamma[\vf]\equiv \Gamma[\m,\vf]$ also depends on
the scale $\m$ used to define the renormalized vertex functions.

A quite different object is the Wilsonian effective action $S_{\rm W}[\m,\vf]$. It is
defined as $\Gamma[\m,\vf]$, except that all loop-momenta are only integrated down to $\m$
which serves as an infra-red cutoff.  In theories with
massive particles only, there is no big difference between $S_{\rm W}[\m,\vf]$ and
$\Gamma[\m,\vf]$ (as long as $\m$ is less than the smallest mass). When massless particles are present, as is the case for gauge
theories, the situation is  different. In particular, in supersymmetric gauge
theories there is the so-called Konishi anomaly which can be viewed as an IR-effect.
Although $S_{\rm W}[\m,\vf]$ depends holomorphically on $\m$, this is not the case for
$\Gamma[\m,\vf]$ due to this anomaly.

\section{The $SU(2)$ case, moduli space}

What Seiberg and Witten achieved, and what will occupy the rest of these notes, is to
determine the Wilsonian effective action in the case where the microscopic theory one
starts with is the $SU(2)$, $N=2$ super Yang-Mills theory \dxii\ or \dxvii. As
explained above (see \dxiv), classically this theory has a scalar potential $V(\f)=\half \tr
([\f^+,\f])^2$. Unbroken susy requires that $V(\f)=0$ in the vacuum, but this still
leaves the possibilities of non-vanishing $\f$ with $[\f^+,\f]=0$.
We are interested in determining the gauge inequivalent vacua. 
A general $\f$ is of the form $\f(x)=\half\sum_{j=1}^3  \left( a_j(x)+i b_j(x)\right)
\sigma_j$ with real fields $a_j(x)$ and $b_j(x)$ (where I assume that not all three $a_j$
vanish, otherwise exchange the roles of the $a_j$'s and $b_j$'s in the sequel). By a
$SU(2)$ gauge transformation one can always arrange $a_1(x)=a_2(x)=0$.
Then $[\f, \f^+]=0$ implies
$b_1(x)=b_2(x)=0$ and hence, with $a= a_3 + i b_3$, one has $\f=\half a \sigma_3$.
Obviously, in the vacuum $a$ must be a constant.
Gauge transformation from the Weyl group (i.e. rotations by
$\pi$ around the 1- or 2-axis of $SU(2)$) can still change $a\to -a$, so $a$ and $-a$ are gauge
equivalent, too. The gauge invariant quantity describing inequivalent vacua is $\half
a^2$, or $\tr\f^2$, which
 is the same, semiclassically.
When quantum fluctuations are important this is no longer so.
In the sequel, we will use the following definitions for $a$ and $u$:
$$u=\la \tr\f^2\ra\quad , \quad \la\f\ra = \half a\sigma_3\ .
\eqn\ti$$
The complex parameter $u$ labels gauge inequivalent vacua. The manifold of gauge
inequivalent vacua is called the moduli space $\M$ of the theory. Hence $u$ is a
coordinate on $\M$, and $\M$ is essentially the complex $u$-plane. We will see in the
sequel that $\M$ has certain singularities, and the knowledge of the behaviour of the
theory near the singularities will eventually allow the determination of the effective
action $S_{\rm W}$.

Clearly, for non-vanishing $\la\f\ra$, the $SU(2)$ gauge symmetry is broken by the
Higgs mechanism, since the $\f$-kinetic term $\vert\nm\f\vert^2$ generates masses for
the gauge fields. With the above conventions, $A_\m^b,\ b=1,2$ become massive with
masses given by $\half m^2 = {1\over g^2}\vert g a\vert^2$, i.e $m=\rd a$. Similarly
due to the $\f,\l,\p$ interaction terms, $\p^b, \l^b,\ b=1,2$ become massive with the
same mass as the $A_\m^b$, as required by supersymmetry. Obviously, $A_\m^3,\p^3$ and
$\l^3$, as well as the mode of $\f$ describing the flucuation of $\f$ in the
$\sigma_3$-direction, remain massless. These massless modes are described by a
Wilsonian low-energy effective action which has to be $N=2$ supersymmetry invariant,
since, although the gauge symmetry is broken, $SU(2)\to U(1)$, the $N=2$ susy remains
unbroken. Thus it must be of the general form \dxviii\ or \dxx\ where the indices $a,b$
now take only a single value ($a,b=3$) and will be suppressed since the gauge
group is $U(1)$. Also, $V$ in \dxx\ is in the adjoint representation and it is easy to
see that from $e^{-2gV}=1-2gV+\ldots$ only the 1 can contribute. In other words, in an
abelian theory there is no self-coupling of the gauge boson and the same arguments
extend to all members of the $N=2$ susy multiplet: they do not carry electric charge.
Thus for a $U(1)$-gauge theory, from \dxx\ we get simply 
$${1\over 16\pi} \im\ix \left[ \int\dt \cf''(\F)W^\a W_\a
+\int\dt\dtb \F^+\cf'(\F) \right] \ . 
\eqn\tii$$

\section{Metric on moduli space}

Consider the second term of the effective action \tii.
In component fields this term reads
$${1\over 4\pi}\im\ix\left[ \cf''(\f)\vert\d_\m\f\vert^2-i\cf''(\f)\p\sm\d_\m\pb
+\ldots\right]
\eqn\tiii$$
where $+\ldots$ stands for non-derivative terms. Similarly, the first term in \tii\
gives
$${1\over 4\pi}\im\ix\left[ \cf''(\f) \left(-{1\over 4}\right) \fmn(F^{\m\n}-i\tilde
F^{\m\n})
-i\cf''(\f)\l\sm\d_\m\lb
+\ldots\right] \ .
\eqn\tiv$$
If we think of these kinetic terms as a four dimensional sigma-model, then the
$\cf''(\f)$ or rather $\im \cf''(\f)$ that appears for all of them plays the role of a
metric in field space. By the same token it defines the metric in the space of
(inequivalent) vacuum
configurations, i.e. the metric on moduli space. From the $\f$-kinetic term one sees
that a sensible definition of the metric on the moduli space is ($\bar a$ denotes the
complex conjugate of $a$)
$$\rmd s^2=\im \cf''(a) \rmd a \rmd \bar a = \im \tau(a) \rmd a \rmd \bar a
\eqn\tv$$
where $\tau(a)=\cf''(a)$ is the effective (complexified) coupling constant in analogy
with \dxix. The $\s$-model metric $G_{\f\f^+}\sim \im\cf''(\f)$ has been replaced on
the moduli space $\M$ by its expectation value in the vacuum corresponding to the
given point on $\M$, i.e. by $\im \cf''(a)=\im\tau(a)$.

The question now is whether the description of the effective action in terms of the
fields $\F, W$ and the function $\cf$ is appropriate for all vacua, i.e. for all value
of $u$, i.e. on all of moduli space. In particular the kinetic terms \tiii, \tiv, or
what is the same, the metric \tv\ on moduli space should be positive definite,
translating into $\im \tau(a) >0$. However, a simple argument shows that this cannot be
the case: since $\cf(a)$ is holomorphic, $\im\tau(a)=\im {\d^2\cf(a)\over \d a^2}$ is a
harmonic function and as such it cannot have a minimum, and hence (on the compactified
complex plane)  it cannot obey
$\im\tau(a)>0$ everywhere (unless it is a constant as in the classical case). The way
out is to allow for different local descriptions: the coordinates $a, \bar a$ and the
function $\cf(a)$ are appropriate  only in a certain region of $\M$. When a singular
point with $\im\tau(a)\to 0$ is approached one has to use a different set of
coordinates $\hat a$ in which $\im\hat\tau (\hat a)$ is non-singular (and
non-vanishing). This is possible provided the singularity of the metric \tv\ is only a
coordinate singularity, i.e. the kinetic terms of the effective action are not
intrinsically singular, which will be the case.

\section{Asymptotic freedom and the one-loop formula}

Classically the function $\cf$ is given by $\half\tau_{\rm class}\P^2$. The one-loop
contribution has been determined in [\SEI]. The combined tree-level and one-loop result
is
$$\cf_{\rm pert}(\P)={i\over 2\pi}\P^2\ln {\P^2\over \L^2} \ .
\eqn\tvi$$
Here $\L^2$ is some combination of $\m^2$ and numerical factors chosen so as to fix the
normalisation of $\cf_{\rm pert}$. Note that due to non-renormalisation theorems for
$N=2$ susy there are no corrections from two or more loops to the Wilsonian effective
action $S_{\rm W}$ and \tvi\ is the full perturbative result. There are however
non-perturbative corrections that will be determined below.

For very large $a$ the dominant contribution when computing $S_{\rm W}$ from the
microscopic $SU(2)$
 gauge theory comes from regions of large momenta ($p\sim a$) where the microscopic
theory is asymptotically free. Thus, as $a\to\infty$ the effective coupling constant
goes to zero, and the perturbative expression \tvi\ for $\cf$ becomes an excellent
approximation. Also $u\sim \half a^2$ in this limit.\foot{
One can check from the explicit solution in section 6 that one indeed has 
$\half a^2 - u = {\cal O}(1/u)$ as $u\to\infty$.
}
Thus
$$\eqalign{
\cf(a  )&\sim {i\over 2\pi} a^2\ln {a^2\over \L^2} \cr
\tau(a) &\sim {i\over \pi} \left( \ln {a^2\over \L^2} +3\right) \cr} \quad {\rm as}\
u\to\infty \ .
\eqn\tvii$$
Note that due to the logarithm appearing at one-loop, $\tau(a)$ is a multi-valued
function of $a^2\sim 2u$. Its imaginary part, however, $\im\tau(a)\sim {1\over
\pi}\ln{\vert a\vert^2\over \L^2}$ is single-valued and positive (for $a^2\to \infty$).

{\bf \chapter{Duality}}
\sectionnumber=0

As already noted, $a$ and $\bar a$ do provide local coordinates on the moduli space
$\M$ for the region of large $u$. This means that in this region $\F$ (and $\F^+$) and
$W^\a$ are appropriate fields to describe the low-energy effective action. As also
noted, this description cannot be valid globally, since $\im\cf''(a)$, being a harmonic
function, must vanish somewhere, unless it is a constant - which it is not. Duality will
provide a different set of (dual) fields $\Fd$ and $W^\a_{\rm D}$ that provide an
appropriate description for a different region of the moduli space.

\section{Duality transformation}

Consider the form \tii\ of the effective action. Define a field dual to $\F$ by
$$\Fd=\cf'(\F)
\eqn\qi$$
and a function $\cfd(\Fd)$ dual to $\cf(\F)$ by
$$\cfd'(\Fd)=-\F
\eqn\qii$$
where, of course, $\cfd'(\Fd)$ means $\rmd \cfd(\Fd)/\rmd \Fd$.  These duality
transformations simply constitute a Legendre transformation\foot{
This was pointed out to me by Frank Ferrari.}
$\cfd(\Fd)=\cf(\F)-\F\Fd$ with $\Fd$ defined as in \qi. Equation \qii\ then is the
standard inverse relation that follows from the Legendre transform. Using these
relations, the second term in the action \tii\ can be written as
$$\eqalign{
\im\ix\dt\dtb\F^+\cf'(\F)&=\im\ix\dt\dtb \left(-\cfd'(\Fd)\right)^+ \Fd \cr
&=
\im\ix\dt\dtb  \Fd^+ \cfd'(\Fd) \ .\cr}
\eqn\qiii$$
We see that this second term in the effective action \tii\ is invariant under the
duality transformation \qi, \qii.

The reader will recognise the similarity of \qi, \qii\ with a canonical transformation.
Indeed $\cf'(\F)=\d\cf/\d\F$ ressembles a (complex) 
momentum (remember that the effective action
is $\sim \ix\dt\dtt\cf(\F)$, eq. \dxviii), so that the second term in \tii\ is like 
$\im\int q^* p = {i\over 2}\int (p^*q-q^*p)$ 
and the duality transformation is $q_D=p$ and $p_D=-q$ which clearly is a
canonical transformation. It is well-known that canonical transformations preserve the
phase-space measure. As a consequence, if the functional integral is formulated  as a
phase-space integral ($\sim\int {\cal D}\F{\cal D}\Pi \exp[\int \dot \F\Pi-H]$),
under appropriate conditions, the Jacobian for the integration measure is unity for
canonical transformations. The present duality transformation is a particularly simple
canonical transformation and we expect the Jacobian to be one.

Next, consider the $\cf''(\F)W^\a W_\a$-term in the effective action \tii. While the
duality transformation \qi, \qii\ on $\F$ is local, this will not be the case for the
transformation of $W^\a$. Recall that $W$ contains the $U(1)$ field strength $\fmn$,
cf. eq. \dvi. This $\fmn$ is not arbitrary but of the form $\d_\m A_\n-\d_\n A_\m$ for
some $A_\m$. This can be translated into the Bianchi identity $\half\e^{\m\n\rho\sigma}
\d_\n F_{\rho\sigma}\equiv \d_\nu \tilde F^{\m\n}=0$. The corresponding constraint in
superspace is $\im (D_\a W^\a)=0$ where $D_\a$ is the same superspace derivative as in
\dvii. This constraint is a consequence of the abelian version of the expression \dvii\
of $W$ in terms of $V$. In the functional integral one has the choice of integrating
over $V$ only, or over $W^\a$ and imposing the constraint $\im (D_\a W^\a)=0$ by a real
Lagrange multiplier superfield which we call $V_D$:
$$\eqalign{
&\int{\cal D}V \exp\left[ {i\over 16\pi}\im\ix\dt\cf''(\F)W^\a W_\a\right]\cr
&\simeq \int{\cal D}W {\cal D}V_D
 \exp\left[ {i\over 16\pi}\im\ix\left(\int\dt \cf''(\F)W^\a W_\a
+{1\over 2}\int\dt\dtb V_D D_\a W^\a\right) \right] \ .\cr}
\eqn\qiv$$
Observe that 
$$\eqalign{
\int\dt\dtb V_D D_\a W^\a&= - \int\dt\dtb D_\a V_D W^\a
=+\int\dt \bar D^2 (D_\a V_D W^\a)\cr
&=\int\dt (\bar D^2 D_\a V_D) W^\a
=-4\int\dt (W_D)_\a W^\a\cr}
\eqn\qiva$$
 where we used $\bar D_{\dot\b} W^\a=0$ and where the dual $W_D$ is defined from $V_D$
 in analogy with the abelian
version of \dvii\ as $(W_D)_\a=-{1\over 4} \bar D^2 D_\a V_D$. Then one can do the
functional integral over $W$ and one obtains
$$\int{\cal D}V_D \exp\left[ {i\over 16\pi}\im\ix\dt
\left( -{1\over\cf''(\F)} W_D^\a W_{D\a}\right)\right] \ .
\eqn\qv$$

This reexpresses the ($N=1$) supersymmetrized  Yang-Mills action in terms of a dual
Yang-Mills action with the effective coupling $\tau(a)=\cf''(a)$ replaced by $-{1\over
\tau(a)}$. Recall that $\tau(a)={\theta(a)\over 2\pi}+{4\pi i \over g^2(a)}$, so that
$\tau\to -{1\over \tau}$ generalizes the inversion of the coupling constant discussed
in the introduction. Also, it can be shown that $W_D$ actually describes the
electromagnetic dual $\fmn\to \tilde F_{\m\n}$, so that the manipulations
leading to \qv\ constitute a  duality transformation that 
generalizes the old electromagnetic duality of Montonen and Olive (cf.
the introduction). Expressing the $-{1\over \cf''(\F)}$ in terms of $\Fd$ one sees from
\qii\ that $\cfd''(\Fd)=-{\rmd \F\over\rmd\Fd}=-{1\over\cf''(\F)}$ so that
$$-{1\over \tau(a)}=\tau_D(\ad) \ .
\eqn\qvi$$
The whole action \tii\ can then equivalently be written as
$${1\over 16\pi} \im\ix \left[ \int\dt \cfd''(\Fd)W_D^\a W_{D\a}
+\int\dt\dtb \Fd^+\cfd'(\Fd) \right] \ . 
\eqn\qvii$$

\section{The duality group}

To discuss the full group of duality transformations of the action it is most
convenient to write it as
$${1\over 16\pi} \im \ix\dt {\rmd\Fd\over \rmd \F}W^\a W_\a +
{1\over 32 i \pi} \ix\dt\dtb \left( \F^+\Fd-\Fd^+\F\right) \ .
\eqn\qvii$$
While we have shown in the previous subsection that there is a duality symmetry
$$\pmatrix{\Fd\cr \F\cr} \to \pmatrix{0&1\cr -1&0\cr} \pmatrix{\Fd\cr \F\cr} \ ,
\eqn\qviii$$
the form \qvii\ shows that there  also is a symmetry
$$\pmatrix{\Fd\cr \F\cr} \to \pmatrix{1&b\cr 0&1\cr} \pmatrix{\Fd\cr \F\cr} \quad ,
\quad b\in {\bf Z} \ .
\eqn\qix$$
Indeed, the second term in \qvii\ remains invariant since $b$ is real, while the first term in
\qvii\ gets shifted by
$${b\over 16\pi} \im\ix\dt W^\a W_\a = - {b\over 16\pi}  \ix \fmn\tilde F^{\m\n} =-2\pi
b \n
\eqn\qx$$
where $\n\in {\bf Z}$ is the instanton number. Since the action appears as $e^{iS}$ in
the functional integral, two actions differing only by $2\pi  {\bf Z}$ are equivalent,
and we conclude that \qix\ with integer $b$ is a symmetry of the effective action. The
transformations \qviii\ and \qix\ together generate the group $Sl(2,{\bf Z})$. This is
the group of duality symmetries.

Note that the metric \tv\ on moduli space can be written as
$$\rmd s^2 =\im(\rmd \ad\rmd \bar a) = {i\over 2} (\rmd a\rmd\bar \ad - \rmd \ad \rmd
\bar a)
\eqn\qxa$$
where $\la\fd\ra = \half \ad\sigma_3$ and $\ad=\d\cf(a)/\d a$, and that this metric
obviously also is invariant under the duality group  $Sl(2,{\bf Z})$

\section{Monopoles, dyons and the BPS mass spectrum}

At this point, I will have to add a couple of ingredients without much further
justification and refer the reader to the literature for more details.

In a spontaneously broken gauge theory as the one we are considering, typically there
are solitons (static, finite-energy solutions of the equations of motion) that carry
magnetic charge and behave like non-singular magnetic monopoles [\THO] (for a
pedagogical treatment, see [\COL]). The duality transformation \qviii\ constructed above
exchanges electric and magnetic degrees of freedom, hence electrically charged states,
as would be described by hypermultiplets of our $N=2$ supersymmetric version, with
magnetic monopoles.

In $N=2$ susy theories there are two types of multiplets: small (or short) ones (4
helicity states) and large (or long) ones (16 helicity states). Massless states must be 
in short
multiplets, while massive states are in short ones if they satisfy $m^2=2\vert
Z\vert^2$, $Z$ being the central charge of the $N=2$ susy algebra, or in long ones if
$m^2>2\vert Z\vert^2$ [\OW]. The states that become massive by the Higgs mechanism must
be in short multiplets since they were before the symmetry breaking (if one imagines
turning on the scalar field expectation value), and the Higgs mechanism cannot generate
the missing $16-4=12$ helicity states. For purely electrically charged states one has
$Z=a n_e$ where $n_e$ is the (integer) electric charge. Duality then implies that a
purely magnetically charged state has $Z=\ad n_m$ where $n_m$ is the (integer)
magnetic charge. A state with both types of charge, called a dyon,
 has $Z=a n_e + \ad n_m$ since the
central charge is additive. All this applies to states in short multiplets, so-called
BPS-states. The mass formula  for these states then is
$$m^2=2\vert Z\vert^2\quad , \quad Z= (n_m,n_e)\pmatrix{\ad\cr a\cr} \ .
\eqn\qxi$$
It is clear that under a $Sl(2,{\bf Z})$ transformation
$M=\pmatrix{\a&\b\cr\g&\delta\cr}
\in Sl(2,{\bf Z})$ acting on $\pmatrix{\ad\cr a\cr}$, the charge vector gets
transformed to $ (n_m,n_e) M =  (n'_m,n'_e)$ which are again integer charges. In
particular, one sees again at the level of the charges that
 the transformation \qviii\ exchanges purely electrically charged states with purely
magnetically charged ones. It can be shown [\SEN,\OLI,\SW] that precisely those BPS
states are stable for which $n_m$ and $n_e$ are  relatively prime, i.e. for stable
states $(n_m,n_e) \ne (qm,qn)$ for integer $m,n$ and $q\ne \pm 1$.

{\bf \chapter{Singularities and Monodromy}}
\sectionnumber=0

In this section we will study the behaviour of $a(u)$ and $\ad(u)$ as $u$ varies on the
moduli space $\M$. Particularly useful information will be obtained from their
behaviour as $u$ is taken around a closed contour. If the contour does not encircle
certain singular points to be determined below, $a(u)$ and $\ad(u)$ will return to
their initial values once $u$ has completed its contour. However, if the $u$-contour
goes around these singular points, $a(u)$ and $\ad(u)$ do not return to their initial
values but rather to certain linear combinations thereof: one has a non-trivial
monodromy for the multi-valued functions $a(u)$ and $\ad(u)$.

\section{The monodromy at infinity}

This is immediately clear from the  behaviour near $u=\infty$. As already
explained in section 3.4, as $u\to \infty$, due to asymptotic freedom,
the perturbative expression for $\cf(a)$ is valid and one has from \tvi\
for $\ad=\d\cf(a)/\d a$
$$\ad(u)={i\over \pi} a \left( \ln{a^2\over \L^2}+1\right)\quad , \quad
u\to\infty \ .
\eqn\ci$$
Now take $u$ around a counterclockwise contour of very large radius in
the complex $u$-plane, often simply written as $u\to e^{2\pi i}u$. This
is equivalent to having $u$ encircle the point at $\infty$ on the Riemann
sphere in a {\it clockwise} sense. In any case, since $u=\half a^2$ (for
$u\to\infty$) one has $a\to -a$ and
$$\ad\to {i\over\pi} (-a) \left( \ln{e^{2\pi i}a^2\over \L^2}+1\right)
=-\ad+2a
\eqn\cii$$
or
$$\pmatrix{\ad(u)\cr a(u)\cr}\to M_\infty  \pmatrix{\ad(u)\cr a(u)\cr}
\quad , \quad M_\infty=\pmatrix{-1&2\cr 0&-1\cr} \ .
\eqn\ciii$$
Clearly, $u=\infty$ is a branch point of $\ad(u)\sim {i\over
\pi}\sqrt{2u} \left(\ln{u\over\L^2}+1\right)$. This is why this point is referred to as a
singularity of the moduli space.

\section{How many singularities?}

Can $u=\infty$ be the only singular point? Since a branch cut has to
start and end somewhere, there must be at least one other singular point.
Following Seiberg and Witten, I will argue that one actually needs three
singular points at least. To see why two cannot work, let's suppose for a
moment that there are only two singularities and show that this leads to
a contradiction.

Before doing so, let me note that there is an important 
so-called $U(1)_R$-symmetry in the classical theory
that takes $\f\to e^{2i\a}\f$, $W \to e^{i\a}W$, $\t\to e^{i\a}\t$,
$\tb\to e^{i\a}\tb$, thus $\dt \to e^{-2i\a} \dt$,
$\dtb \to e^{-2i\a}\dtb$ and hence $\P \to e^{2i\a}\P$, so that the
classical action \dxii\ or \dxvii\ is invariant under this global
symmetry. More generallly, the action \dxviii\ will be invariant if
$\cf(\P) \to e^{4i\a} \cf(\P)$. This symmetry is broken by the one-loop
correction and also by instanton contributions. The latter give
corrections to $\cf$ of the form $\P^2\sum_{k=1}^\infty c_k \left( \L^2/
\P^2 \right)^{2k}$, and hence are invariant only for 
$\left(e^{4i\a}\right)^{2k} =1$, i.e. $\a={2\pi n\over 8},\ n\in{\bf Z}$.
Hence instantons break the $U(1)_R$-symmetry  to a dicrete ${\bf Z}_8$.
The one-loop corrections behave as 
${i\over 2\pi}\P^2\ln {\P^2\over\L^2}\to e^{4i\a}\left(
{i\over 2\pi}\P^2\ln {\P^2\over\L^2} - {2\a\over \pi}\P^2\right)$. As 
in the paragraph before eq. \qx\ one shows that this only changes the
action by $2\pi\n \left({4\a\over \pi}\right)$ where $\n$ is integer, so
that again this change is irrelevant as long as ${4\a\over \pi}=n$ or
$\a={2\pi n\over 8}$. Under this ${\bf Z}_8$-symmetry, $\f\to e^{i\pi
n/2}\f$, i.e. for odd $n$ one has $\f^2\to -\f^2$. The non-vanishing
expectation value $u=\la\tr\f^2\ra$ breaks this ${\bf Z}_8$ further to
${\bf Z}_4$. Hence for a given vacuum, i.e. a given point on moduli space
there is only a ${\bf Z}_4$-symmetry left from the $U(1)_R$-symmetry.
However, on the manifold of all possible vacua, i.e. on $\M$, one has
still the full ${\bf Z}_8$-symmetry, taking $u$ to $-u$.

Due to this global symmetry $u\to -u$, singularities of $\M$ should come
in pairs: for each singularity at $u=u_0$ there is another one at
$u=-u_0$. The only fixed points of $u\to -u$ are $u=\infty$ and $u=0$. We
have already seen that $u=\infty$ is a singular point of $\M$. So if
there are only two singularities the other must be the fixed point $u=0$.

If there are only two singularities, at $u=\infty$ and $u=0$, then by contour
deformation (``pulling the contour over the back of the sphere")\foot{
It is well-known from complex analysis that monodromies are associated with contours
around branch points. The precise from of the contour does not matter, and it can be
deformed as long as it does not meet another branch point. Our singularities precisely
are the branch points of $a(u)$ or $\ad(u)$.}
 one sees that the
monodromy around 0 (in a counterclockwise sense) is the same as the above monodromy
around $\infty$: $M_0=M_\infty$. But then $a^2$ is not affected by any monodromy and
hence is a good global coordinate, so
one can take $u=\half a^2$ on all of $\M$, and furthermore one must have
$$\eqalign{\ad&={i\over \pi} a \left( \ln {a^2\over \L^2}+1\right) + g(a)\cr
a&=\sqrt{2u}\cr}
\eqn\civ$$
where $g(a)$ is some entire function of $a^2$. This implies that
$$\tau={\rmd\ad\over\rmd a}={i\over \pi} \left( \ln {a^2\over \L^2}+3\right) +
{\rmd g\over\rmd a} \ .
\eqn\cv$$
The function $g$ being entire, $\im {\rmd g\over\rmd a}$ cannot have a minimum (unless
constant) and it is clear that $\im\tau$ cannot be positive everywhere. As already
emphasized, this means that $a$ (or rather $a^2$) cannot be a good global coordinate and
\civ\ cannot hold globally. Hence, two singularities only cannot work.

The next simplest choice is to try 3 singularities. Due to the $u\to -u$ symmetry,
these 3 singularities are at $\infty, u_0$ and $-u_0$ for some $u_0\ne 0$. 
In particular, $u=0$ is no
longer a singularity of the quantum moduli space. To get a singularity also at $u=0$
one would need at least four singularities at $\infty, u_0, -u_0$ and $0$. As discussed
later, this is not possible, and more generally, exactly 3 singularities seems to be
the only consistent possibility.

So there is no singularity at $u=0$ in the quantum moduli space $\M$.
Classically, however, one precisely expects that $u=0$ should be a singular point,
since classically $u=\half a^2$, hence $a=0$ at this point, and
then there is no Higgs mechanism any more. Thus all (elementary) massive states, i.e.
the gauge
bosons $A_\m^1, A_\m^2$ and their susy partners $\p^1, \p^2, \l^1, \l^2$
become massless. Thus the description of the lights fields in terms of
the previous Wilsonian effective action should break down, inducing a singularity on
the moduli space. As already stressed, this is the clasical picture. While $a\to
\infty$ leads to asymptotic freedom  and the microscopic $SU(2)$ theory is weakly
coupled, as $a\to 0$ one goes to a strong coupling regime where the classical
reasoning has no validity any more, and $u\ne \half a^2$. By the BPS mass formula \qxi\
massless gauge bosons still are possible at $a=0$, but this does no longer correspond to
$u=0$.

So where has the singularity due to massless gauge bosons at $a=0$ 
moved to? One might be tempted to think that
$a=0$ now corresponds to the singularities at $u=\pm u_0$, but this is not the case
as I will show in a moment. The answer is that the point $a=0$ no
longer belongs to the quantum moduli space (at least not to the component connected to
$u=\infty$ which is the only thing one considers). This can be seen explicitly from
the form of the solution for $a(u)$ given in the next section.

\section{The strong coupling singularities}

Let's now concentrate on the case of three singularities at $u=\infty, u_0$ and $-u_0$.
What is the interpretation of the (strong-coupling) singularities at finite $u=\pm
u_0$? One might first try to consider that they are still due to the gauge bosons
becoming massless. However, as Seiberg and Witten point out, massless gauge bosons
would imply an asymptotically conformally invariant theory in the infrared limit and
conformal invariance implies $u=\la\tr\f^2\ra=0$ unless $\tr\f^2$ has dimension zero
and hence would be the unity operator - which it is not. So the singularities at $u=\pm
u_0\ (\ne 0)$ do not correspond to massless gauge bosons.

There are no other elementary $N=2$ multiplets in our theory. The next thing to try
is to consider collective excitations - solitons, like the magnetic monopoles or dyons.
Let's first study what happens if a magnetic monopole  of unit magnetic charge becomes
massless. From the BPS mass formula \qxi, the mass of the magnetic monopole is
$$m^2=2\vert \ad\vert^2
\eqn\cvi$$
and hence vanishes at $\ad=0$. We will see that this produces one of the two
stron-coupling singularities. So call $u_0$ the value of $u$ at whiche $\ad$ vanishes.
Magnetic monopoles are described by hypermultiplets $M$ of $N=2$ susy that couple
locally to the dual fields $\Fd$ and $W_D$, just as electrically charged ``electrons"
would be described by hypermultiplets that couple locally to $\F$ and $W$. So in the
dual description we have $\Fd, W_D$ and $M$, and, near $u_0$, $\ad\sim \la \Fd\ra$ is
small. This theory is exactly $N=2$ susy QED with very light electrons (and a subscript
$D$ on every quantity). The latter theory is not asymptotically free, but has a
$\b$-function given by
$$\m{\rmd\over \rmd\m} g_D={g_D^3\over 8\pi}
\eqn\cvii$$
where $g_D$ is the coupling constant. But the scale $\m$ is proportional to $\ad$ and
${4\pi i\over g_D^2(\ad)}$ is $\tau_D$ for $\t_D=0$ (of course, super QED, unless
embedded into a larger gauge group,  does not
allow for a non-vanishing theta angle). One concludes that for $u\approx u_0$ or
$\ad\approx 0$
$$\ad {\rmd\over \rmd \ad} \tau_D=-{i\over \pi} \ \Rightarrow \ \tau_D = -{i\over \pi}
\ln\ad \ .
\eqn\cviii$$
Since $\tau_D={\rmd (-a)\over \rmd\ad}$ this can be integrated to give
$$a\approx a_0+{i\over \pi} \ad\ln\ad \qquad (u\approx u_0)
\eqn\cix$$
where we dropped a subleading term $-{i\over \pi}\ad$. Now, $\ad$ should be a good
coordinate in the vicinity of $u_0$, hence depend linearly\foot{
One might want to try a more general dependence like $\ad\approx c_0 (u-u_0)^k$ with
$k>0$. This leads to a monodromy in $Sl(2,{\bf Z})$ only for integer $k$. The factorisation
condition below, together with the form of $M(n_m,n_e)$ also given below, then imply that
$k=1$ is the only possibility.
}
 on $u$. One concludes
$$\eqalign{\ad&\approx c_0 (u-u_0)\cr
a&\approx  a_0+{i\over \pi} c_0 (u-u_0)\ln (u-u_0) \ .\cr}
\eqn\cx$$
From these expressions one immediately reads the monodromy as $u$ turns around $u_0$
counterclockwise, $u-u_0\to e^{2\pi i} (u-u_0)$:
$$
\pmatrix{\ad\cr a\cr} \to \pmatrix{ \ad\cr a-2\ad \cr} = M_{u_0} 
\pmatrix{\ad\cr a\cr} \quad , \quad
 M_{u_0}=\pmatrix{1&0\cr -2&1\cr} \ . 
\eqn\cxi$$

To obtain the monodromy matrix at $u=-u_0$ it is enough to observe that the contour
around $u=\infty$ is equivalent to a counterclockwise contour of very large radius in
the complex plane. This contour can be deformed into a contour encircling $u_0$ and a
contour encircling $-u_0$, both counterclockwise. It follows the factorisation
condition on the monodromy matrices\foot{
There is an ambiguity concerning the ordering of $M_{u_0}$ and $M_{-u_0}$
which will be resolved below.}
$$M_\infty=M_{u_0} M_{-u_0}
\eqn\cxii$$
and hence
$$M_{-u_0} = \pmatrix{-1&2\cr -2&3\cr} \ .
\eqn\cxiii$$

What is the interpretation of this singularity at $u=-u_0$? To discover this, consider
the behaviour under monodromy of the BPS mass formula $m^2=2\vert Z\vert^2$ with $Z$
given by \qxi, i.e. $Z=(n_m,n_e)\pmatrix{\ad\cr a\cr}$. The monodromy transformation
$\pmatrix{\ad\cr a\cr}\to M \pmatrix{\ad\cr a\cr}$ can be interpreted as changing the
magnetic and electric quantum numbers as
$$(n_m,n_e)\to (n_m,n_e) M \ .
\eqn\xiv$$
The state of vanishing mass responsible for a singularity should be invariant under the
monodromy, and hence be a left eigenvector of $M$ with unit eigenvalue. This is clearly
so for the magnetic monopole: $(1,0)$ is a left eigenvector of $\pmatrix{1&0\cr
-2&1\cr}$ with unit eigenvalue. This simply reflects that $m^2=2\vert \ad\vert^2$ is
invariant under \cxi. Similarly, the left eigenvector of \cxiii\ with unit eigenvalue
is $(n_m, n_e)=(1,-1)$ This is a dyon. Thus the sigularity at $-u_0$ is interpreted as
being due to a $(1,-1)$ dyon becoming massless.

More generally, $(n_m, n_e)$ is the left eigenvector with unit eigenvalue\foot{
Of course, the same is true for any $(q n_m, q n_e)$ with $q\in {\bf Z}$, but according
to the discussion in section 4.3 on the stability of BPS states, states with $q\ne \pm
1$ are not stable.}
of
$$M(n_m,n_e)=\pmatrix{ 1+2n_m n_e& 2 n_e^2\cr -2 n_m^2 & 1- 2 n_m n_e\cr}
\eqn\cxv$$
which is the monodromy matrix that should appear for any singularity due to a massless
dyon with charges $(n_m, n_e)$. Note that $M_\infty$ as given in \ciii\ is not of this
form, since it does not correspond to a hypermultiplet becoming massless.

One notices that the relation \cxii\ does not look invariant under $u\to -u$, i.e
$u_0\to -u_0$ since $M_{u_0}$ and $M_{-u_0}$ do not commute. The apparent contradiction
with the ${\bf Z}_2$-symmetry is resolved by the following remark. The precise
definition of the composition of two monodromies as in \cxii\ requires a choice of
base-point $u=P$ (just as in the definition of homotopy groups). Using a different
base-point, namely $u=-P$, leads to 
$$M_\infty =M_{-u_0}M_{u_0}
\eqn\cxvi$$
instead. Then one would obtain $M_{-u_0}=\pmatrix{3&2\cr -2&-1}$, and comparing with
\cxv, this would be interpreted as due to a $(1,1)$ dyon. Thus the ${\bf
Z}_2$-symmetry $u\to -u$ on the quantum moduli space also acts on the base-point $P$,
hence exchanging \cxii\ and \cxvi. At the same time it exchanges the $(1,-1)$ dyon
with the $(1,1)$ dyon.

Does this mean that the $(1,1)$ and  $(1,-1)$ dyons play a privileged role? Actually
not. If one first turns $k$ times around $\infty$, then around $u_0$, and then $k$
times around $\infty$ in the opposite sense, the corresponding monodromy is
$M_\infty^{-k} M_{u_0} M_\infty^k = \pmatrix{1-4k&8k^2\cr -2& 1+4k\cr}=M(1,-2k)$ and
similarly
$M_\infty^{-k} M_{-u_0} M_\infty^k =
 \pmatrix{-1-4k&2+8k+8k^2\cr -2& 3+4k\cr} =
M(1,-1-2k)$. So one sees that these monodromies correspond to dyons with $n_m=1$ and any
$n_e\in {\bf Z}$ becoming massless. Similarly one has e.g.
$M_{u_0}^{k} M_{-u_0} M_{u_0}^{-k}=M(1-2k,-1)$, etc.

Let's come back to the question of how many singularities there are. Suppose there are
$p$ strong coupling singularities at $u_1, u_2, \ldots u_p$ in addition to the one-loop
perturbative singularity at  $u=\infty$. Then one has a factorisation analogous to
\cxii:
$$M_\infty =  M_{u_1}  M_{u_2} \ldots  M_{u_p}
\eqn\cxvii$$
with $M_{u_i}=M(n_m^{(i)}, n_e^{(i)})$ of the form \cxv. It thus becomes a problem of
number theory to find out whether, for given $p$, there exist solutions to  \cxvii\ with integer 
$n_m^{(i)}$ and $n_e^{(i)}$. For several low values of $p>2$ it has been 
checked [\LER]
that there are no such solutions, and it seems likely that the same is true for all
$p>2$.

{\bf \chapter{The solution : determination of the low-energy effective action}}
\sectionnumber=0

So far we have seen that $\ad(u)$ and $a(u)$ are single-valued except for the
monodromies around $\infty, u_0$ and $-u_0$. As is well-known from complex analysis,
this means that  $\ad(u)$ and $a(u)$ are really multi-valued functions with branch
cuts, the branch points being  $\infty, u_0$ and $-u_0$. A typical example is 
$f(u)=\sqrt{u} F(a,b,c;u)$, where $F$ is the hypergeometric function. The latter has a
branch cut from $1$ to $\infty$. Similarly, $\sqrt{u}$ has a branch cut from $0$ to
$\infty$ (usually taken along the negative real axis), so that $f(u)$ has two branch
cuts joining the three singular points $0,1$ and $\infty$. When $u$ goes around any of
these singular points there is a non-trivial monodromy between $f(u)$ and one other
function $g(u)= u^d F(a',b',c';u)$. The three monodromy matrices are in (almost) one-to-one
correspondence with the pair of functions $f(u)$ and $g(u)$.

In the physical problem at hand one knows the monodromies, namely
$$M_{\infty}=\pmatrix{-1&2\cr 0&-1\cr}\ , \quad 
M_{u_0}=\pmatrix{1&0\cr -2&1\cr}\ , \quad
M_{-u_0}=\pmatrix{-1&2\cr -2&3\cr}
\eqn\si$$
and one wants to determine the corresponding functions $\ad(u)$ and $a(u)$. As will be
explained, the monodromies fix $\ad(u)$ and $a(u)$ up to normalisation, which will be
determined from the known asymptotics \ci\ at infinity.

The precise location of $u_0$ depends on the renormalisation
conditions which can be chosen such that $u_0=1$ [\SW]. Assuming this choice in the
sequel will simplify somewhat the equations. If one wants to keep $u_0$, essentially
all one has to do is to replace $u\pm 1$ by ${u\pm u_0\over u_0}={u\over u_0}\pm 1$.

\section{The differential equation approach}

Monodromies typically arise from differential equations with periodic coefficients.
This is well-known in solid-state physics where one considers a Schr\"odinger equation
with a periodic potential\foot{
The constant energy has been included into the potential, and the mass has been
normalised to $\half$.}
$$\left[ -{\rmd^2\over \rmd x^2} + V(x)\right] \p(x)=0 \quad , \quad V(x+2\pi)=V(x) \ .
\eqn\sia$$
There are two independent solutions $\p_1(x)$ and $\p_2(x)$. One wants to compare
solutions at $x$ and at $x+2\pi$. Since, due to the periodicity of the potential $V$,
the differential equation at $x+2\pi$ is exactly the same as at $x$, the set of
solutions must be the same. In other words, $\p_1(x+2\pi)$ and $\p_2(x+2\pi)$ must be
linear combinations of $\p_1(x)$ and $\p_2(x)$:
$$\pmatrix{\p_1\cr \p_2\cr} (x+2\pi) = M \pmatrix{\p_1\cr \p_2\cr} (x) 
\eqn\sii$$
where $M$ is a (constant) monodromy matrix.

The same situation arises for differential equations in the complex plane with
meromorphic coefficients. Consider again the Schr\"odinger-type equation
$$\left[ -{\rmd^2\over \rmd z^2} + V(z)\right] \p(z)=0
\eqn\siii$$
with meromorphic $V(z)$, having poles at $z_1, \ldots z_p$ and (in general) also at
$\infty$. The periodicity of the previous example is now replaced by the
single-valuedness of $V(z)$ as $z$ goes around any of the poles of $V$ (with $z-z_i$
corresponding roughly to $e^{ix}$). So, as $z$  goes once around any one of the $z_i$, the
differential equation \siii\  does not change. So by the same argument as above, the two solutions
$\p_1(z)$ and $\p_2(z)$, when continued along the path surrounding $z_i$ must again be
linear combinations of $\p_1(z)$ and $\p_2(z)$:
$$\pmatrix{\p_1\cr \p_2\cr} \left(z+e^{2\pi i}(z-z_i)\right) 
= M_i \pmatrix{\p_1\cr \p_2\cr} (z) 
\eqn\siv$$
with a constant $2\times 2$-monodromy matrix $M_i$ for each of the poles of $V$. Of
course, one again has the factorisation condition \cxvii\ for $M_\infty$. It is
well-known, that non-trivial constant monodromies correspond to poles of $V$ that are
at most of second order. In the language of differential equations, \siii\ then only
has {\it regular} singular points.

In our physical problem, the {\it two} multivalued functions $\ad(z)$ and $a(z)$ have 3
singularities with non-trivial monodromies at $-1, +1$ and $\infty$. Hence they must be
solutions of a second-order differential equation \siii\ with the potential $V$ having
(at most) second-order poles precisely at these points. 
The general form of this potential is\foot{
Additional terms in $V$ that naively look like first-order poles 
($\sim {1\over z-1}$ or ${1\over z+1}$)
cannot appear since they correspond to third-order poles at $z=\infty$.
}
$$V(z)=-{1\over 4} \left[ {1-\l_1^2\over (z+1)^2} +  {1-\l_2^2\over (z-1)^2 }
-{1-\l_1^2-\l_2^2+\l_3^2\over (z+1)(z-1)} \right]
\eqn\sv$$
with double poles at $-1, +1$ and $\infty$. The corresponding residues are 
$-{1\over 4}(1-\l_1^2)$, $-{1\over 4}(1-\l_2^2)$ and $-{1\over 4}(1-\l_3^2)$. Without
loss of generality, I assume $\l_i\ge 0$. The corresponding differential equation
\siii\ is well-known in the mathematical literature (see e.g. [\ERD]) 
since it can be transformed into the hypergeometric differential equation. It has
appeared, among others, in the study of the (classical) Liouville three-point function
and the determination of constant curvature metrics on Riemann surfaces [\BG]. The
transformation to the standard hypergeometric equation is readily performed by setting
$$\p(z)=(z+1)^{\half (1-\l_1)} (z-1)^{\half (1-\l_2)}\,  f\left( {z+1\over 2}\right) \ .
\eqn\svi$$
One then finds that $f$ satisfies the hypergeometric differential equation
$$x(1-x) f''(x)+[c-(a+b+1)x]f'(x)-abf(x)=0
\eqn\sviii$$
with
$$\eqalign{
a&=\half (1-\l_1-\l_2+\l_3)\cr
b&=\half (1-\l_1-\l_2-\l_3)\cr
c&=1-\l_1 \ . \cr }
\eqn\six$$
The solutions of the hypergeometric equation \sviii\ can be written in many different
 ways due to the various identities between the hypergeometric function $F(a,b,c;x)$
and products with powers, e.g. $(1-x)^{c-a-b} F(c-a,c-b,c;x)$, etc. A convenient choice for the two
independent solutions is the following [\ERD]
$$\eqalign{
f_1(x)&=(-x)^{-a}F(a,a+1-c,a+1-b;{1\over x})\cr
f_2(x)&=(1-x)^{c-a-b} F(c-a,c-b,c+1-a-b;1-x) \ . \cr }
\eqn\sx$$
$f_1$ and $f_2$ correspond to Kummer's solutions denoted $u_3$ and $u_6$. The choice of
$f_1$ and $f_2$ is motivated by the fact that $f_1$ has simple monodromy properties
 around
$x=\infty$ (i.e. $z=\infty$) and $f_2$ has simple monodromy properties
around $x=1$ (i.e. $z=1$),
so they are good candidates to be identified with $a(z)$ and $\ad(z)$.

One can extract a great deal of information from the asymptotic forms of $\ad(z)$ and
$a(z)$. As $z\to\infty$ one has $V(z)\sim -{1\over 4} \, {1-\l_3^2\over z^2}$, so that
the two independent solutions behave asymptotically as $z^{\half (1\pm \l_3)}$ if $\l_3
\ne 0$, and as $\sqrt{z}$ and $\sqrt{z}\ln z$ if $\l_3=0$. Comparing with \civ\ 
(with $u\to z$) we see
that the latter case is realised. Similarly, with $\l_3=0$, as $z\to 1$, one has
$V(z)\sim -{1\over 4} \left(  {1-\l_2^2\over (z-1)^2}-{1-\l_1^2-\l_2^2\over 2(z-1)}
\right)$, where I have kept the subleading term. From the logarithmic asymptotics \cx\
one then concludes $\l_2=1$ (and from the subleading term also $-{\l_1^2\over
8}={i\over \pi}{c_0\over a_0}$). The ${\bf Z}_2$-symmetry ($z\to -z$) on the moduli
space then implies that, as $z\to -1$, the potential  $V$ does not  have a double pole
either, so that also $\l_1=1$. Hence we conclude
$$\l_1=\l_2=1\ ,\ \ \l_3=0 \ \Rightarrow \ V(z)= -{1\over 4}\, {1\over (z+1)(z-1)}
\eqn\sxi$$
and $a=b=-\half,\ c=0$. Thus from \svi\ one has $\p_{1,2}(z)=f_{1,2}\left({z+1\over
2}\right)$. One can then verify, using the formulas in ref. [\ERD] (and denoting the
argument again by $u$ rather than $z$) that the two solutions
$$\eqalign{
\ad(u)&=i \p_2(u)=i{u-1\over 2} F\left({1\over 2},{1\over 2},2;{1-u\over 2}\right)\cr
a(u)&=-2i \p_1(u)=\sqrt{2} (u+1)^{1\over 2} F\left(-{1\over 2},{1\over 2},1;{2\over u+1}\right)\cr }
\eqn\sxiii$$
indeed have the required monodromies \si, as well as the correct asymptotics.

It might look as if we have not used the monodromy properties to determine $\ad$ and
$a$ and that they have been determined only from the asymptotics. This is not entirely
true, of course. The very fact that there are non-trivial monodromies only at $\infty,
+1$ and $-1$ implied that $\ad$ and $a$ must satisfy the second-order differential
equation \siii\ with the potential \sv. To determine the $\l_i$ we then used the
asymptotics of $\ad$ and $a$. But this is (almost) the same as using the monodromies
since the
latter were obtained from the asymptotics.

Using the integral representation [\ERD] of the hypergeometric function, the solution
\sxiii\ can be nicely rewritten as 
$$\eqalign{
\ad(u)&={\sqrt{2}\over \pi} \int_1^u {\rmd x\ \sqrt{x-u}\over \sqrt{x^2-1} } \cr
a(u)&={\sqrt{2}\over \pi} \int_{-1}^1 {\rmd x\ \sqrt{x-u}\over \sqrt{x^2-1} } \ . \cr }
\eqn\sxiv$$

One can invert the second equation \sxiii\  to obtain $u(a)$ and 
insert the result into $\ad(u)$ to obtain $\ad(a)$. Integrating
with respect to $a$ yields $\cf(a)$ and hence the low-energy effective action. I should
stress that this expression for $\cf(a)$ is not globally
valid but only on a certain portion of the moduli space. Different analytic
continuations must be used on other portions.

\section{The approach using elliptic curves}

In their paper, Seiberg and Witten do not use the differential equation approach just
described, but rather introduce an auxiliary construction: a certain elliptic curve by
means of which two functions with the correct monodromy properties are constructed. I
will not go into details here, but simply sketch this approach.

To motivate their construction {\it a posteriori}, we notice the following: from the
integral representation \sxiv\ it is natural to consider the complex $x$-plane. More
precisely, the integrand has square-root branch cuts with branch points at $+1, -1, u$
and $\infty$. The two branch cuts can be taken to run from $-1$ to $+1$ and from $u$ to
$\infty$. The Riemann surface of the integrand is two-sheeted with the two sheets
connected through the cuts. If one adds the point at infinity to each of the two
sheets, the topology of the Riemann surface is that of two spheres connected by two
tubes (the cuts), i.e. a torus. So one sees that the Riemann
surface of the integrand in \sxiv\ has genus one. This is the elliptic curve considered
by Seiberg and Witten.

As is well-known, on a torus there are two independent non-trivial closed paths
(cycles). One cycle ($\g_2$) can be taken to go once around the cut $(-1,1)$, and the
other cycle ($\g_1$) to go from $1$ to $u$ on the first sheet and back from $u$ to $1$
on the second sheet. The solutions $\ad(u)$ and $a(u)$ in \sxiv\ are precisely the
integrals of some suitable differential $\l$ along the two cycles $\g_1$ and $\g_2$:
$$\ad=\oint_{\g_1} \l  \quad , \quad a=\oint_{\g_2} \l  \quad , \quad
\l={\sqrt{2}\over 2\pi} {\sqrt{x-u}\over \sqrt{x^2-1} } \rmd x \ .
\eqn\sxiva$$
These integrals are called period integrals. They are known to satisfy a
second-order differential equation, the so-called Picard-Fuchs equation, that is
nothing else than our Schr\"odinger-type equation \siii\ with $V$ given by \sxi.

How do the monodromies appear in this formalism? As $u$ goes once around $+1,-1$ or
$\infty$, the cycles $\g_1, \g_2$ are changed into linear combinations of themselves
with integer coefficients:
$$\pmatrix{\g_1\cr \g_2} \to M \pmatrix{\g_1\cr \g_2} \quad , \quad
M\in Sl(2, {\bf Z}) \ .
\eqn\sxv$$
This immediately implies
$$\pmatrix{\ad\cr a} \to M \pmatrix{\ad\cr a}
\eqn\sxvi$$
with the same $M$ as in \sxv. The advantage here is that one automatically gets
monodromies with {\it integer} coefficients. The other advantage is that
$$\tau(u)={\rmd \ad/\rmd u\over  \rmd a/\rmd u}
\eqn\sxvii$$
can be easily seen to be the $\tau$-parameter describing the complex structure of the
torus, and as such is garanteed to satisfy
$$\im\tau(u) >0
\eqn\sxviii$$
which was the requirement for positivity of the metric on moduli space.

To motivate the appearance of the genus-one elliptic curve (i.e. the torus) {\it a
priori} - without knowing the solution \sxiv\ from the differential equation approach -
Seiberg and Witten remark that the three monodromies are all very special: they do not
generate all of $Sl(2, {\bf Z})$ but only a certain subgroup $\G(2)$ of matrices 
in
$Sl(2, {\bf Z})$ congruent to $1$ modulo $2$. Furthermore, they remark that the
$u$-plane with punctures at $1,-1,\infty$ can be thought of as the  quotient of the
upper half plane $H$ by $\G(2)$, and that $H/\G(2)$ naturally parametrizes (i.e. is the
moduli space of) elliptic curves described by
$$y^2=(x^2-1)(x-u) \ .
\eqn\sxix$$
Equation \sxix\ corresponds to the genus-one Riemann surface discussed above, and it is
then natural to introduce the cycles $\g_1, \g_2$ and the differential $\l$ from \sxiv.
The rest of the argument then goes as I just exposed.

{\bf \chapter{Conclusions and outlook}}
\sectionnumber=0

In these notes, I have given a rather detailed, and hopefully pedagogical introduction
to the work of Seiberg and Witten [\SW]. We have seen realised a version of
electric-magnetic duality accompanied by a duality transformation on the expectation
value of the scalar (Higgs) field, $a\leftrightarrow \ad$. There is a manifold of
inequivalent vacua, the moduli space $\M$, corresponding to different Higgs expectation
values. The duality relates strong coupling regions in $\M$ to the perturbative region
of large $a$ where the effective low-energy action is known asymptotically
 in terms of $\cf$. Thus duality allows us to determine the latter also at strong
coupling. The holomorphicity condition from $N=2$ supersymmetry then puts such strong
constraints on $\cf(a)$, or equivalently on $\ad(u)$ and $a(u)$ that the full functions
can be determined solely from their asymptotic behaviour at the strong and weak
coupling singularities of $\M$.

There are a couple of questions one might ask, like what is the profound reason for the
appearance of elliptic curves, or of the differential equation. It is intriguing to
note that the latter  with the potential \sv\ appears in conformal field
theories as the null vector decoupling equation. It is satisfied by certain chiral conformal
four-point correlation functions
$$\la {\cal V}_{\Delta_4}(\infty)  {\cal V}_{\Delta_3}(1)  {\cal V}_{\Delta_2}(z)
 {\cal V}_{\Delta_1}(0) \ra
$$
where the ${\cal V}_{\Delta}$ are chiral vertex operators and where 
the conformal dimensions $\Delta_j$ are determined in terms of the $\l_1, \l_2,
\l_3$. Whether this is a pure coincidence or has some deeper meaning does not seem to
be clear at the moment.

Also, several generalisations of the pure $SU(2)$ Yang-Mills theory exposed here
 have been studied. One is to add matter hypermultiplets [\MAT], another 
is to consider
pure Yang-Mills theory but for gauge groups different from $SU(2)$ [\LER,\OGG], 
or to allow for
different gauge groups as well as matter [\DS]. Here let me only note that for the pure
$SU(3)$ theory, solving the condition $[\f, \f^+]=0$ leads to $\f=a_1 H_1 + a_2 H_2$
where  $H_i$ are the two Cartan generators of $SU(3)$, so that one has a two-complex
dimensional moduli space, parametrized by $a_1, a_2$ or rather by $u=\la\tr\f^2\ra$ and
$v=\la\tr\f^3\ra$. The duals are $a_{Di}={\d\cf\over \d a_i},\ i=1,2$. The monodromies
in moduli space (i.e. the $(u,v)$-space) then act on the four-component object 
$(a_{D1}(u,v),a_{D2}(u,v),a_1(u,v),a_2(u,v))$. They can be reproduced
 from period integrals of some hyperelliptic curve [\LER]. The corresponding
(Picard-Fuchs) differential equations are two-partial differential equations in $u$ and
$v$  [\LER] with solutions given by Appel functions [\ERD] that generalise the
hypergeometric function to two variables.

Last, but not least, I should mention that similar duality ideas in string theory have
led to yet another explosion of this domain of theoretical physics. A particular nice
link with the field theory discussed here has been made in [\MVL] where the field
theoretic duality is related to string dualities.

\ack

It is a pleasure to thank all participants of the duality study group at the Ecole
Normale for discussions, Luis Alvarez-Gaum\'e for making his notes [\ALG] available to
me well before publication, and Marc Rosso for giving me the opportunity to present
this lecture at the Strasbourg meeting.

\refout

\end